\documentclass[aps,prl,twocolumn,amsmath,amssymb,reprint,numbers,superscriptaddress,noeprint,longbibliography]{revtex4-2}

\usepackage{braket}
\usepackage{soul}
\usepackage{geometry}
\usepackage{graphicx}
\usepackage{booktabs}
\usepackage{array}
\usepackage{relsize}
\usepackage{color}
\usepackage{xcolor}
\usepackage{eqnarray}
\usepackage{caption}
\usepackage{textcase}
\usepackage{textcomp}
\usepackage{setspace}
\usepackage{xcolor}
\usepackage{siunitx}
\usepackage{placeins}
\usepackage[normalem]{ulem}

\usepackage{style}
\myexternaldocument{build}{supplement}

\usepackage{subfig}
\usepackage{tikz}
\usetikzlibrary{positioning}
\hypersetup{
     colorlinks   = true,
     linkcolor    = blue,
     citecolor    = blue,
     urlcolor = blue
}

\definecolor{cola}{rgb}{0.7,0.1,0.1}
\definecolor{colb}{rgb}{0.9,0.4,0}
\definecolor{colc}{rgb}{0.3,0.7,0}
\definecolor{cold}{rgb}{0,0.35,0.75}
\definecolor{cole}{rgb}{0.63, 0.13, 0.94}
\definecolor{colf}{rgb}{0.5, 0.5, 0.5}

\newcommand{\cmV}{cm\textsuperscript{3}}
\newcommand{\cmnV}{cm\textsuperscript{-3}}
\newcommand{\degC}{\textsuperscript{o}C}

\newcommand{\Do}{D\textsuperscript{0}}
\newcommand{\DoX}{D\textsuperscript{0}X}

\newcommand{\Yo}{Y\textsubscript{0}}

\newcommand{\InZnX}{In$_\mathrm{Zn}^\mathrm{0}$X}

\newcommand{\AlX}{Al$^\mathrm{0}$X}
\newcommand{\GaX}{Ga$^\mathrm{0}$X}
\newcommand{\InX}{In$^\mathrm{0}$X}

\newcommand{\InZn}{In$_\mathrm{Zn}^\mathrm{0}$}

\newcommand{\In}{In$^\mathrm{0}$}

\newcommand{\transInX}{\InZn\,$\leftrightarrow$\,\InZnX}

\newcommand{\Otwo}{O\textsubscript{2}}
\newcommand{\Xep}{Xe\textsuperscript{+}}

\newcommand{\Bparc}{$\vec{B} \parallel \hat{c}$}
\newcommand{\Bperpc}{$\vec{B} \perp \hat{c}$}

\begin{document}

\title{Isolation of Single Donors in ZnO}

\author{Ethan R. Hansen}
\thanks{E. R. Hansen and V. Niaouris contributed equally to this work. Correspond at: ethanrh@uw.edu, niaouris@uw.edu.} 
\affiliation{Department of Physics, University of Washington, Seattle, WA, 98195, USA}
\author{Vasileios Niaouris}
\thanks{E. R. Hansen and V. Niaouris contributed equally to this work. Correspond at: ethanrh@uw.edu, niaouris@uw.edu.} 
\affiliation{Department of Physics, University of Washington, Seattle, WA, 98195, USA}
\author{Bethany E. Matthews}
\affiliation{Energy and Environment Directorate, Paciﬁc Northwest National Laboratory, Richland, WA, 99352, USA}
\author{Christian Zimmermann}
\affiliation{Department of Physics, University of Washington, Seattle, WA, 98195, USA}
\author{Xingyi Wang}
\affiliation{Department of Electrical and Computer Engineering, University of Washington, Seattle, WA, 98195, USA}
\author{Roman Kolodka}
\affiliation{Department of Physics, University of Washington, Seattle, WA, 98195, USA}
\author{Lasse Vines}
\affiliation{Department of Physics and Centre for Materials Science and Nanotechnology, University of Oslo, Blindern,
N-0316 Oslo, Norway}
\author{Steven R. Spurgeon}
\affiliation{Department of Physics, University of Washington, Seattle, WA, 98195, USA}
\affiliation{National Security Directorate, Paciﬁc Northwest National Laboratory, Richland, WA, 99352, USA}
\author{Kai-Mei C. Fu}
\affiliation{Department of Physics, University of Washington, Seattle, WA, 98195, USA}
\affiliation{Department of Electrical and Computer Engineering, University of Washington, Seattle, WA, 98195, USA}
\affiliation{Physical Sciences Division, Pacific Northwest National Laboratory, Richland, WA, 99352, USA}

\begin{abstract}
The shallow donor in zinc oxide (ZnO) is a promising semiconductor spin qubit with optical access. Single indium donors are isolated in a commercial ZnO substrate using plasma focused ion beam (PFIB) milling. Quantum emitters are identified optically by spatial and frequency filtering. The indium donor assignment is based on the optical bound exciton transition energy and magnetic dependence. The single donor emission is intensity and frequency stable with a transition linewidth less than twice the lifetime limit. The isolation of optically stable single donors post-FIB fabrication is promising for optical device integration required for scalable quantum technologies based on single donors in direct band gap semiconductors.
\end{abstract}

\date{\today}

\maketitle

Optically accessible solid-state defects have favorable properties for photon-based applications in quantum computing~\cite{benjamin2009qce, weber2010qcd, ladd2010qc} and quantum communication~\cite{wehner2018qiv,orieux2016rai}.
ZnO, a direct band gap II-VI semiconductor, is a promising host material for spin qubits with optical access. 
ZnO combines the efficient electron-photon coupling present in direct band gap III-V semiconductors~\cite{he2013its,degreve2012qds,gao2012oeb} with low spin-orbit coupling~\cite{niaouris2022esr} and the potential of nuclear spin-free lattice, characteristics both present in silicon and diamond which have led to long spin-coherence times~\cite{tyryshkin2012esc, balasubramanian2009usc}.
Prior works studying neutral shallow donor ensembles in natural isotopic abundance ZnO, specifically ensembles of Al, Ga, and In substituting for Zn, have shown promising optical and spin properties, including ensemble optical linewidths narrow enough to enable opticalS spin manipulation~\cite{wagner2011bez,linpeng2018cps,wang2023pdq, niaouris2023col}, longitudinal spin lifetimes up to 0.5\,s~\cite{niaouris2022esr} and coherence times up to 50~\textmu s which are limited by substrate purity~\cite{linpeng2018cps}.
These emitters exhibit an estimate Huang-Rhys factor of $0.06$~\cite{wagner2011bez} and oscillator strength of $0.35$ (Supplemental Material~\cite{supp}\nocite{dexter1958top, hilborn1982ecc, meyer2004bed, rossler1999iic, chen2013ddb, nichelatti2002aed, linpeng2020dqd, lambrecht2002vbo, wagner2009vbs} Sec.~\ref*{si:osc_strength}).
Recent work on implanted In ensembles~\cite{wang2023pdq} has also revealed the large, 100\,MHz hyperfine coupling of the In electron spin-$\frac{1}{2}$ to the In nuclear spin-$\frac{9}{2}$~\cite{block1982odm, gonzalez1982mrs, buss2016omm}.
Thus, there is a path toward deterministic formation of In donors with access to a nuclear spin memory~\cite{neumann2008mea, morton2008ssq, merkel2008qch}.

Most quantum information applications require individually addressable qubits. 
The challenge of isolating single donors in ZnO~\cite{viitaniemi2022csp, wang2023pdq} stems from the low chemical purity of available substrates compared to host materials like silicon and diamond. 
This difficulty is compounded by the ease of forming single atom substitutional point defects compared to defect complexes such as the nitrogen-vacancy and silicon-vacancy center in diamond~\cite{smith2019ccg} and radiation damage centers in silicon~\cite{dreau2021bdn}. 
Commercially available ZnO crystals, with total donor concentrations exceeding $10^{16}$\,\cmnV, render optical isolation of single donors unattainable through optical confocal imaging. 
An alternative is to reduce the material volume and focus on lower density impurity species, such as In. 
While single donors in the direct band gap semiconductor ZnSe have been isolated in very narrow quantum wells~\cite{karasahin2022sqe}, there exists a high-level of inhomogeneity induced by the quantum-well potential. 
Here we seek solutions that maintain the bulk donor and donor-bound exciton properties. 
Focused ion beam (FIB) milling has been extensively used to extract and shape materials at the nanoscale~\cite{melngailis1987fib, matsui1996fib, tseng2005rdn,moll2018fib}, 
enabling the fabrication efficient photon extraction devices ~\cite{,deluca2016fib,ho2007fib,manoccio2021fib} for quantum defects~\cite{zhong2018oas}. 
FIB fabrication, however, results in a damaged layer of material ~\cite{rubanov2004fid,rubanov2005dcd} which can degrade the optical and spin properties of some quantum defects ~\cite{bayn2011dpf,sarcan2023uih}. 
Given that ZnO is a radiation resilient material~\cite{kucheyev2002eiz}, FIB milling may be a viable method for isolating single In donors without diminishing their favorable spin and optical properties.

We demonstrate the optical isolation of single In donors in ZnO via \Xep\ plasma FIB (PFIB) milling. 
Following PFIB processing, the optical properties of the ZnO are severely degraded; however, a 1-hour oxygen anneal recovers sharp donor-bound exciton (\DoX) to donor-bound electron (\Do) photoluminescence (PL) lines. 
Single In donor candidates are optically isolated with spectral and spatial filtering. 
Verification of emitters corresponding to In donors is achieved via collection of the two-electron satellite transition under resonant excitation and magneto-photoluminesescence. 
Lifetime measurements reveal a ten-fold lifetime reduction relative to bulk ZnO, indicative of non-radiative channels. 
Measured linewidths, however, are less than a factor of two broader than the measured lifetime-limit. 
Further materials and fabrication improvements, combined with device integration for Purcell enhancement, could be used to enhance the radiative efficiency.

For this work, a 360\,\textmu m-thick ZnO crystal (Tokyo Denpa) is used as the parent substrate, with the [0001] crystal axis $\hat{c}$ perpendicular to the substrate surface. 
We observe Al, Ga, and In donor-bound exciton (\AlX, \GaX, and \InX) in the substrate. 
The donor concentrations, measured on the back surface, were determined by secondary ion mass spectrometry (SIMS) measurements as $5.3\cdot10^{14}$\,\cmnV\ for Al and $4.4\cdot10^{15}$\,\cmnV\ for Ga (Supplemental Material~\cite{supp} Sec.~\ref*{si:donor_conc}). 
The In concentration was below the SIMS detection limit. 
Since the donor concentration can vary across the substrate, these values provide an order of magnitude estimate.

\begin{figure}[t]
    \centering
    \includegraphics[width=0.99\linewidth]{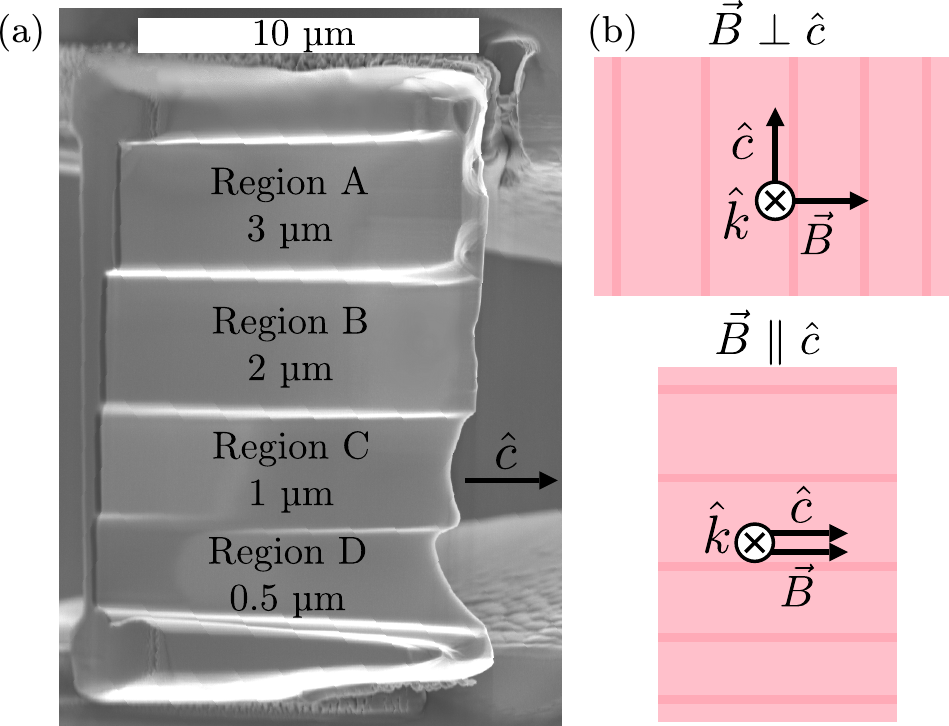}
    \caption{\label{fig:sample}
    (a) Scanning electron microscopy image of lamella.
    (b) Orientation of crystal ($\hat{c}$) and optical ($\hat{k}$) axes with respect to the magnetic field $\vec{B}$.
    }
\end{figure}

Traditionally, FIB technologies use Ga\textsuperscript{+} beams~\cite{tseng2005rdn} which could unintentionally introduce a high-density of Ga donors in ZnO, thus we utilize a PFIB with an inert ion gas source, \Xep, to cut a 5\,\textmu m-thick cross-section (lamella) from the ZnO parent substrate. 
Using standard lift-out techniques, a micro-manipulator needle removed the lamella and laid it over a PFIB-milled trench on a Si\Otwo\ wafer. 
The edges of the lamella were secured with Pt deposited by cracking a metal organic gas (flowed in using a standard gas injection system) with the \Xep-beam.
Tiered steps of thicknesses ranging from 0.5\,\textmu m to 3\,\textmu m were milled using a 30\,keV beam, with the crystal $\hat{c}$ axis parallel to the step edges. 
The lamella was then polished with a 5\,keV beam on both sides to remove re-sputtered material and amorphous damage that often occurs during the higher energy milling processes. 
See Supplemental Material~\cite{supp}, Sec.~\ref*{si:sample_prep} for sample preparation process illustration.
The lamella (Fig.~\ref{fig:sample}a) was annealed at 700\,\degC\ for 1 hour under \Otwo\ flow to reduce remaining milling damage~\cite{wang2023pdq}. 

All measurements were conducted with a confocal microscope that images the lamella inside a helium immersion cryostat with a superconducting magnet (Supplemental Material~\cite{supp} Sec.~\ref*{si:microscope}). 
The magnetic field $\vec{B}$ is always parallel to the lamella plane. 
We access two magnetic field orientations, \Bparc\ and \Bperpc, by rotating the lamella (Fig.~\ref{fig:sample}b).

\begin{figure*}[ht]
    \centering
    \includegraphics[width=\linewidth]{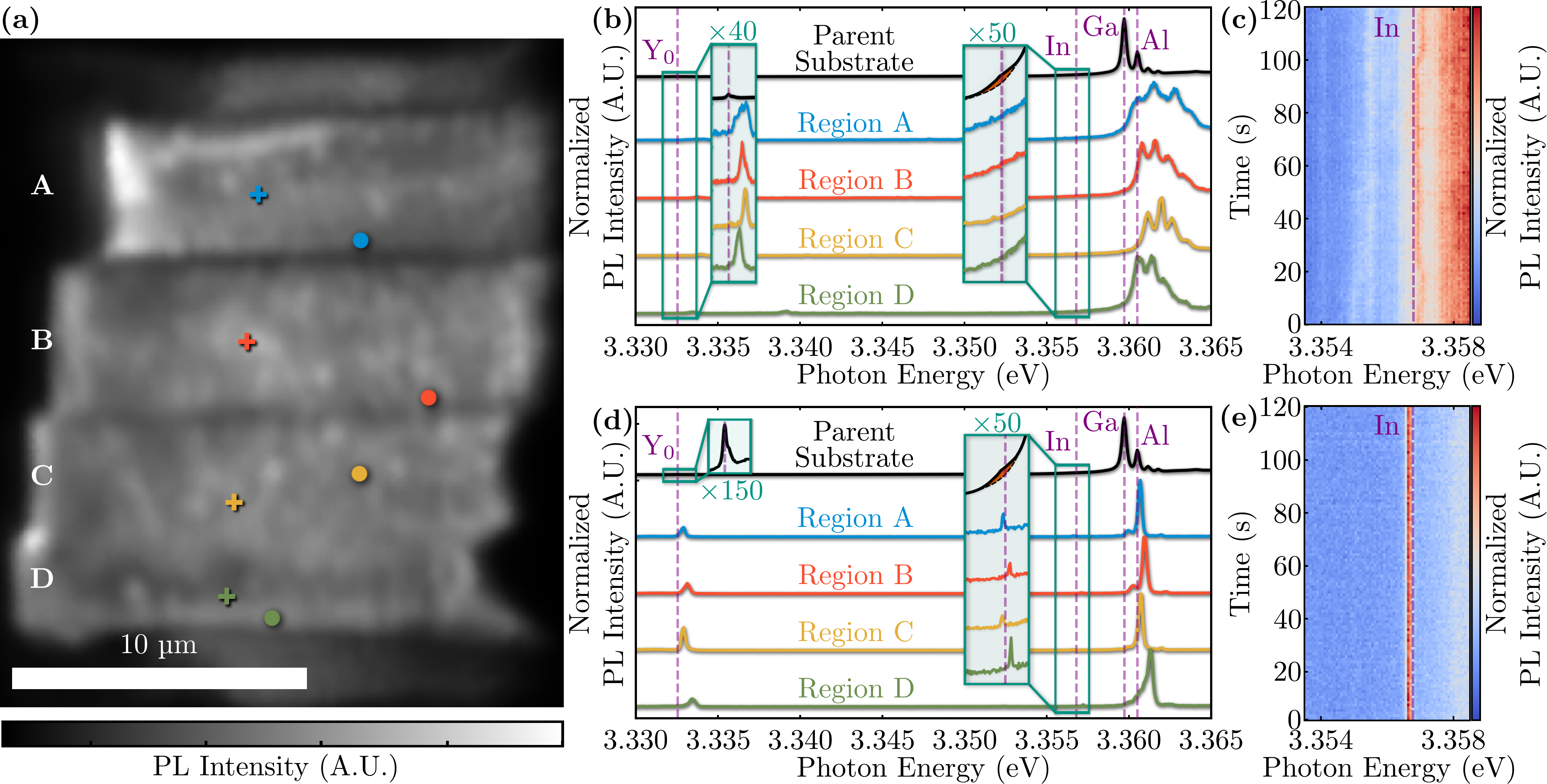}
    \caption{
    Continuous-wave excitation at 3.44\,eV, $\text{T} = 5.2$\,K.
    (a) Confocal PL image of lamella post-anneal. 
    Cross (circle) markers indicate the locations for spectra in Fig.~\ref{fig:spectra}b (Fig.~\ref{fig:spectra}d).
    (b) and (d) PL spectra of pre- and post-annealed lamella respectively, normalized to maximum intensity. 
    Dashed lines mark the \DoX\ transition for different donors~\cite{wagner2011bez}.
    Insets depict expanded views near the \Yo\ (left) and \InX\ (right) lines.
    (c) and (e) PL kinetic series from region C of the pre- and post-annealed lamella, respectively.
    }
    \label{fig:spectra}
\end{figure*}

\begin{figure*}[t]
    \centering
    \includegraphics[width=0.99\linewidth]{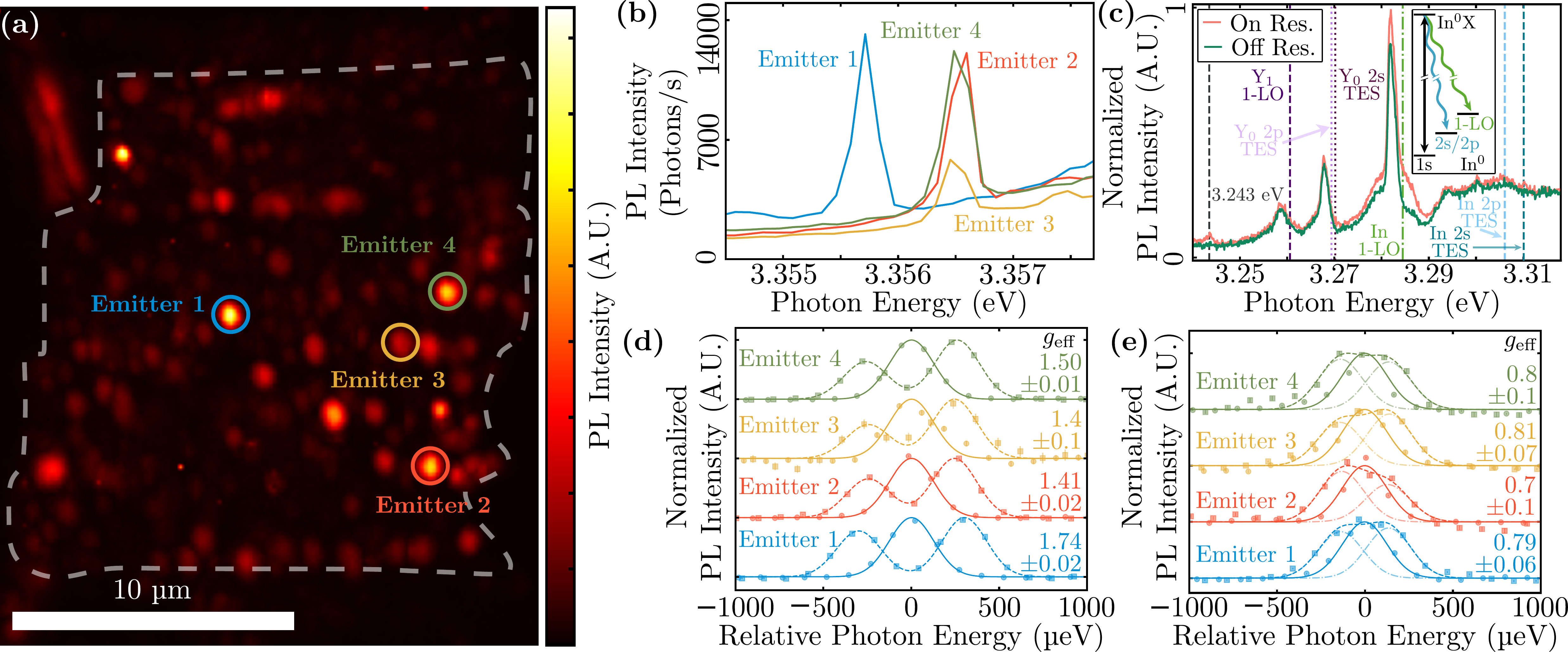}
    \caption{
    Continuous-wave excitation at 3.44\,eV.
    (a) Confocal PL image collecting $3.354-3.357$\,eV. $\text{T} = 5.2$\,K.
    (b) PL spectra of four emitters at  5.2\,K. Spectra are shifted to adjust for strain-induced energy offsets.
    (c) Sideband PL of emitter 2 under resonant excitation at 3.355948\,eV, $\text{T} = 5.2$\,K, normalized to the peak of the on-resonance spectrum.
    Dashed lines mark reported transitions~\cite{wagner2011bez, wang2023pdq}. Inset depicts energy diagram of resonant excitation, sideband PL collection scheme.
    (d) and (e) PL spectra with \Bperpc\ ($\text{T} = 8.2$\,K) and \Bparc\ ($\text{T} = 6.6$\,K), respectively, at 0\,T (circles) and 6\,T (squares), with Gaussian fits. The Al/Ga background emission is subtracted. Spectra are normalized to their maximum value. 
    }
    \label{fig:donor_ver}
\end{figure*}

Fig.~\ref{fig:spectra}a displays a confocal PL image of the lamella post-annealing. 
PL spectra from the substrate and each region of the pre-annealed lamella are shown in Fig.~\ref{fig:spectra}b.
The substrate spectrum shows a small shoulder at the \InX\ transition~\cite{wagner2011bez} with a peak intensity roughly $3\cdot10^{-4}$ times that of the \GaX\ peak. 
Based on the measured Ga density and assuming that the relative peak intensity and donor concentration are proportional, we estimate $10^{12}$ In donors/\cmV. 
With an excitation spot diameter of $\sim$450\,nm, a $\sim$5\,\textmu m thick slice of the substrate would yield one In donor per excitation spot, with smaller concentrations for the $0.5 - 3$\,\textmu m thick steps of the lamella.

A weak \Yo\ line, related to excitons bound to structural defects~\cite{wagner2011bez}, is observed in the substrate PL. 
After PFIB fabrication, an increase in the \Yo\ PL intensity relative to the \GaX\ and \AlX\ lines is observed, which we ascribe to the creation of structural defects during fabrication(Fig.~\ref{fig:spectra}b). 
Additionally, an overall PL shift varying across the lamella is observed, consistent with differential strain environments. 
Within a single excitation spot, the \GaX\ and \AlX\ lines also show splitting and broadening which we attribute to microscopic strain (Fig.~\ref{fig:spectra}b).

We monitored the PL spectra as a function of time in kinetic series measurements; Fig.~\ref{fig:spectra}c shows a series from region C. 
Multiple weak features are observed, with a stronger line at the \InX\ transition. 
Similar features are observed at multiple locations in the lamella (Supplemental Material~\cite{supp}, Sec.~\ref*{si:donor_stability}).
These stronger emitters at the \InX\ transition exhibit discrete spectral jumps, a behavior characteristic of isolated single emitters in an unstable environment~\cite{chakravarthi2021isl, vanDam2019ocd, ourari2023itb}.
 
After annealing, the broad \GaX\ and \AlX\ lines become single, sharp transitions, suggesting that annealing removed a majority of the milling-induced damage (Fig.~\ref{fig:spectra}d).
A small overall PL shift persists after annealing. 
Interestingly, the intensity of the \GaX\ line relative to the \AlX\ line is greatly diminished when compared to the substrate PL. 
The cause of this reduction is unknown.

Near the \InX\ line we now observe spots with strong, spectrometer-resolution limited emission (Fig.~\ref{fig:spectra}d right inset). 
A representative PL kinetic series is shown in Fig.~\ref{fig:spectra}(e). 
Compared to the pre-annealed PL spectrum, the PL emission at the \InX\ transition becomes stronger and spectrally stable.
Additionally, the weaker features around the \InX\ region are no longer observed, suggesting that they may have been surface-related shallow defects which were eliminated via annealing~\cite{feng2021rsd}.

To spatially resolve single In donors, we perform confocal PL scans while spectrally filtering the \InX\ emission; the resulting image is shown in Fig.~\ref{fig:donor_ver}a. 
Localized emission with spectrometer-resolution limited linewidths at \InX\ are observed (Fig.~\ref{fig:donor_ver}b). 
The highest density of emitters is found in regions B and C rather than the thicker region A. We note that the the total PL intensity is also brighter in these regions (Fig.~\ref{fig:spectra}a) and suspect that the PL intensity may be dominated by surface-related non-radiative recombination. 
While 5\,keV PFIB polishing was performed across the entire sample, different regions were polished by different amounts which could account for the variation in PL intensity. 

For the remainder of the paper, we focus on four emitters whose positions are labelled in Fig.~\ref{fig:donor_ver}a. 
The PL spectra corresponding to each emitter is shown in Fig.~\ref{fig:donor_ver}b, where the spectra has been shifted utilizing the location of the \AlX\ and \Yo\ lines to adjust for relative energy offsets induced by strain (Supplemental Material~\cite{supp}, Sec.~\ref*{si:emitter_PL}). 

To verify these emitters as In donors, we resonantly excite the transition between the \In($1s$) state to the \InX\ state and collect the PL sideband consisting of the 1-LO phonon-replica and the two-electron satellites (TES) transitions. 
The TES transitions correspond to relaxation from the lowest \InX\ state to the excited $2s$ and $2p$ states of \In\ (Fig~\ref{fig:donor_ver}c inset). 
Fig~\ref{fig:donor_ver}c displays the sideband PL for emitter 2 with both on- and off-resonant excitation (see Supplemental Material~\cite{supp}, Sec.~\ref*{si:res_tes} for emitters 3 and 4). 
For emitters 2--4, resonant PL enhancement is observed at the In 1-LO, 2s and 2p TES transitions~\cite{wagner2011bez}, as well as an unidentified In-related transition at 3.243\,eV (Supplemental Material~\cite{supp}, Sec.~\ref*{fig:tes_em3_em4_implnt}). 
We observe a large background in the In sideband PL. We believe that this is sideband PL from the tails of the much brighter \AlX\ and \GaX\ zero-phonon transitions upon which the \InX\ transition resides. In contrast with the In-assigned sideband PL, the sideband PL background shifts with laser excitation energy. 
Resonant sideband enhancement is not observed for emitter 1.  

We perform magneto-photoluminescence (magneto-PL) measurements with both \Bperpc\ and \Bparc\ to verify that the observed Zeeman splittings are consistent with emission from neutral donors. 
Fig.~\ref{fig:donor_ver}d (Fig.~\ref{fig:donor_ver}e) shows the PL of each emitter at 0\,T and 6\,T  with \Bperpc\ (\Bparc).
In both orientations, we only collect linearly polarized emission perpendicular to the crystal axis $\hat{c}$, which is expected to be 50 times larger than linearly polarized emission parallel to $\hat{c}$ (Supplemental Material~\cite{supp}, Sec.~\ref*{si:pol_rules} and Refs.~\cite{linpeng2020dqd, lambrecht2002vbo, wagner2009vbs}). 
This strong polarization dependence is also experimentally observed (Supplemental Material~\cite{supp}, Sec.~\ref*{si:pol_rules}).
The effective $g$ factor ($g_{\mathrm{eff}}$) for each orientation is calculated according to $g_{\mathrm{eff}} = \Delta E/\left(\mu_B B\right)$, where $\Delta E$ denotes the energy splitting of bound exciton transitions and $\mu_B$ denotes the Bohr magneton. 
This splitting results from the hole Zeeman splitting of \DoX\ and the electron Zeeman splitting of \Do.
The hole $g$ factor ($g^{\parallel / \perp}_h$) is then calculated according to $g^{\perp}_h = g^{\perp}_{\mathrm{eff}} - g^{\perp}_e$ ($g^{\parallel}_h = g^{\parallel}_e - g^{\parallel}_{\mathrm{eff}}$), where $g^{\perp}_e\simeq g^{\parallel}_e = 1.95$~\cite{wang2023pdq, rodina2004mop}.

In both orientations, the emission of all emitters splits into two peaks linearly dependent on field (Supplemental Material~\cite{supp} Sec.~\ref*{si:magnetoPL}), consistent with a donor. 
If a donor electron is not present, as is the case for an ionized donor-bound exciton, electron-hole coupling yields a non-linear dependence~\cite{rodina2004mop}. 
Measured hole $\left|g\right|$ factors are consistent with ensemble measurements\cite{rodina2004mop, wagner2009vbs,linpeng2018cps}.
With \Bparc, we find $g^{\parallel}_h$ to range between $-1.14$ and $-1.2$ for all emitters. 
With \Bperpc, we find $g^{\perp}_h = 0.22$ for emitter 1, and between $0.46$ and $0.54$ for emitters 2--4, which qualitatively follows the literature where reported $g^{\perp}_h$ values vary between $0.1$ and $0.34$.

\begin{figure}[b]
    \centering
    \includegraphics[width=0.99\linewidth]{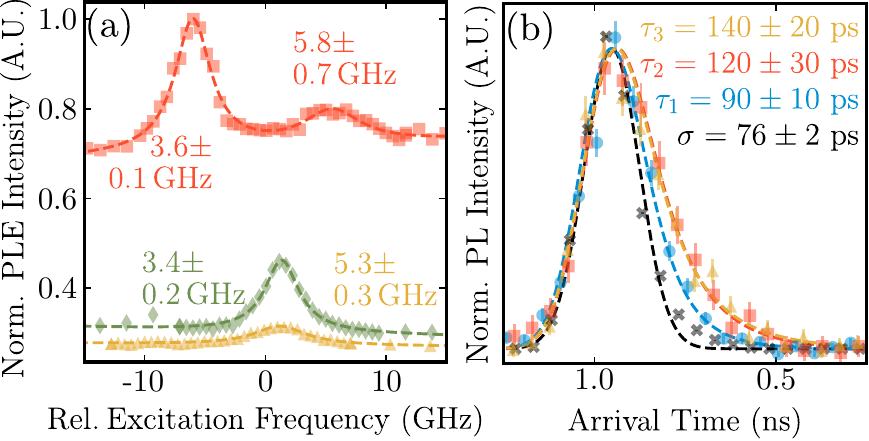}
    \caption{\label{fig:ple_lifetime}
    T = 5.2\,K.
    (a) PLE spectra of emitters 2--4 normalized to the maximum counts of the emitter 2 spectra. Detector dark counts are insignificant.
    Exposure time at each excitation frequency varies between $20-60$\,s, based on emitter intensity. Peaks are fit to Lorenzians and labelled with the fit full-width-at-half-maximum.
    (b) Time-resolved PL of emitters 1--3 under pulsed excitation (2\,ps width, 12.5\,ns repetition rate) at 3.44\,eV. Response function measured with a direct laser reflection and fit to a Gaussian. 
    Time-resolved PL is fit with two exponential-Gaussian convolution profiles to account for the In emitter and the Al/Ga background (Supplemental Material~\cite{supp}, Sec.~\ref*{si:lifetime_bg}). 
    Fits are labelled with the In lifetime.
    }
\end{figure}

We utilize photoluminescence excitation (PLE) spectroscopy to determine the linewidth and spectral stability of emitters 2--4.
In PLE measurements, we resonantly scan a continuous-wave laser over the \transInX\ transition and collect sideband PL (Fig.~\ref{fig:donor_ver}c). 
The spectral stability under resonant excitation, with energy less than the ZnO band gap, supports that the donors remain in the neutral charge state in the absence of semiconductor carrier generation. 
This could be an advantage compared to deep-level defects such as the nitrogen-vacancy center in diamond which undergo two-photon ionization under resonant excitation~\cite{aslam2013pii} and require a charge repump laser~\cite{chakravarthi2021isl, vanDam2019ocd}.

As shown in Fig.~\ref{fig:ple_lifetime}a, the high resolution PLE spectroscopy enables us to determine the absorption linewidth and also resolve multiple emitters. 
For example, for emitter 2 we observe a weak second peak corresponding to a second emitter.
The emitter linewidths range between 3.4\,GHz and 5.8\,GHz.

The expected linewidth of a lifetime-limited emitter at 5.2\,K is estimated by adding three different contributions. 
First, the line is split by 0.5\,GHz due to the zero-field hyperfine interaction between the donor electron spin-$\frac{1}{2}$ and the In nuclear spin-$\frac{9}{2}$ ($A_\text{In}=100$\,MHz)~\cite{gonzalez1982mrs, block1982odm}). 
At 0\,K, the two hyperfine lines would have a lifetime-limited linewidth of 120--150\,MHz, given by the reported In ensemble lifetime~\cite{wagner2011bez,chen2013ddb}.
Each line is further broadened by 0.65\,GHz due to a phonon-assisted thermal population relaxation between \DoX\ states at 5.2\,K~\cite{niaouris2023col}.
Under these effects, the two lines are not resolvable, with a total effective linewidth of 1.2\,GHz; $3$ times smaller than the observed values.
Additionally, the disparity between the expected and measured linewidths could be a result of a shortened lifetime compared to the literature value.

We performed time-resolved PL measurements using pulsed laser excitation to probe the emitter lifetime. 
As shown in Fig.~\ref{fig:ple_lifetime}b,
we observe excited state lifetimes between 90\,ps and 140\,ps, one order of magnitude smaller than expected~\cite{chen2013ddb}. 
Surface-related non-radiative recombination has been suggested to explain the energy-dependent \DoX\ lifetime in bulk ZnO, where high-energy excitation absorbed near the ZnO surface results in shorter measured lifetimes~\cite{chen2013ddb}. 
The intensity of the emission does not increase when lowering the temperature to 2\,K (Supplemental Material~\cite{supp}, Sec.~\ref*{si:temp_dep}), suggesting that the recombination mechanism is not thermally activated.  
The diminished lifetimes in the lamella correspond to an estimated radiative efficiency of $\sim$10\,\% and estimated lifetime-limited linewidths ranging from $1.1-1.8$\,GHz, depending on the emitter. 
The total homogeneous linewidth increases to $2.0-2.6$\,GHz once contributions due to phonons and the hyperfine interaction are included; hence, the measured linewidth, while only $1.3-3$ times larger than the lifetime-limited linewidth, has an additional unidentified broadening component of $1$\,GHz. To realize spin-selective excitation, the non-radiative relaxation and this additional broadening mechanism need to be understood and eliminated.

In summary, we have demonstrated the isolation of single neutral shallow donors in ZnO by employing plasma focused ion beam milling techniques to isolate a small volume of ZnO.
Through resonant excitation and magneto-PL, we identify three emitters as In donors. 
The isolated donors exhibit near lifetime-limited linewidths with charge-stable emission. 
The estimated 10\,\% radiative efficiency motivates future dedicated studies on surface damage and understanding of surface effects on near-surface quantum emitters in ZnO. 
However, ZnO is a fairly new quantum defect host and the relative ease of single donor isolation and observed signal recovery after one annealing step is promising. 
Additionally, the FIB-robust optical signal suggests FIB is appropriate for optical device fabrication in ZnO. 
Compatibility with FIB processing is important due to the monolithic nature of high-purity ZnO and the need for features small/comparable to the 369\,nm transition wavelength. 
Finally, we note that this work further motivates the need for both high purity ZnO material to reduce the band-edge luminescence from unintentional dopants.

\ 

The authors thank Yusuke Kozuka for the bulk ZnO substrates. 
This material is based on work primarily supported by the U.S. Department of Energy (DOE), Office of Science, Office of Basic Energy Sciences, under Award No. DE-SC0020378, and partially supported by the National Science Foundation under Grant No. 1820614 and 2212017. 
B.E.M. and S.R.S. were supported for electron microscopy and interpretation by the DOE Office of Science National Quantum Information Science Research Centers, Co-design Center for Quantum Advantage (C2QA) under contract number DE-SC0012704. 
Pacific Northwest National Laboratory is a multiprogram national laboratory operated for the U.S. Department of Energy (DOE) by Battelle Memorial Institute under Contract No. DE- AC05-76RL0-1830. 
Lifetime measurements were supported by the National Science Foundation DMR-2212017. 
SIMS was supported through project No. 325573, funded by the Norwegian research council.

\bibliography{main.bib}

\end{document}


\title{Supplemental Material for ``Isolation of Single Donors in ZnO''}

\author{Ethan R. Hansen}
\thanks{E. R. Hansen and V. Niaouris contributed equally to this work. Correspond at: ethanrh@uw.edu, niaouris@uw.edu.} 
\affiliation{Department of Physics, University of Washington, Seattle, WA, 98195, USA}
\author{Vasileios Niaouris}
\thanks{E. R. Hansen and V. Niaouris contributed equally to this work. Correspond at: ethanrh@uw.edu, niaouris@uw.edu.} 
\affiliation{Department of Physics, University of Washington, Seattle, WA, 98195, USA}
\author{Bethany E. Matthews}
\affiliation{Energy and Environment Directorate, Paciﬁc Northwest National Laboratory, Richland, WA, 99352, USA}
\author{Christian Zimmermann}
\affiliation{Department of Physics, University of Washington, Seattle, WA, 98195, USA}
\author{Xingyi Wang}
\affiliation{Department of Electrical and Computer Engineering, University of Washington, Seattle, WA, 98195, USA}
\author{Roman Kolodka}
\affiliation{Department of Physics, University of Washington, Seattle, WA, 98195, USA}
\author{Lasse Vines}
\affiliation{Department of Physics and Centre for Materials Science and Nanotechnology, University of Oslo, Blindern,
N-0316 Oslo, Norway}
\author{Steven R. Spurgeon}
\affiliation{Department of Physics, University of Washington, Seattle, WA, 98195, USA}
\affiliation{National Security Directorate, Paciﬁc Northwest National Laboratory, Richland, WA, 99352, USA}
\author{Kai-Mei C. Fu}
\affiliation{Department of Physics, University of Washington, Seattle, WA, 98195, USA}
\affiliation{Department of Electrical Engineering, University of Washington, Seattle, WA, 98195, USA}
\affiliation{Physical Sciences Division, Pacific Northwest National Laboratory, Richland, WA, 99352, USA}

\maketitle
\vspace{-0.4cm}

\section{Oscillator strength and dipole moment of donor bound exciton - donor bound electron transitions in ZnO}
\label{si:osc_strength}

Following Refs.~\cite{dexter1958top, hilborn1982ecc}, we calculate the oscillator strength of the \transAX\ transition from the \transEX\ emission lifetime in the zero-phonon line (ZPL) $\tau_{\text{D,ZPL}}$. 
Specifically, 
\begin{equation}
    \label{eq:osc_strength}
    f_{ab} = \frac{g_b}{g_{a}}\frac{2 \pi \epsilon_0 m^* c^3}{n(\omega_{ba}) e^2 \omega_{ba}^2 \tau_{D,\text{ZPL}}},
\end{equation}
where $a, b$ correspond to the \DoX\ and \Do states respectively, $g_\text{i}$ is the degeneracy of the $i$ state, $\epsilon_0$ is the vacuum electric permittivity, $m^* = 0.27 m_e$ is the effective electron mass in ZnO~\cite{meyer2004bed, rossler1999iic}, $c$ is the speed of light in vacuum, $n(\omega_{ba})$ is the refractive index at the \transEX\ transition frequency, and $\omega_{ba}$ is the \transEX\ transition frequency.
Taking into account the first phonon replica (1LO)~\cite{wagner2011bez} and two electron satellite (TES) emission fractions, as well as the radiative emitter lifetimes for for Al, Ga, and In ($\tau_\text{Al} \sim 0.78 - 0.86$\,ns, $\tau_\text{Ga} \sim 0.80 - 1.06$\,ns, and $\tau_\text{In} \sim 1.05 - 1.35$\,ns~\cite{wagner2011bez, chen2013ddb}), the estimated ZPL lifetimes are $\tau_\text{Al,ZPL} \sim 0.87 - 0.95$\,ns, $\tau_\text{Ga,ZPL} \sim 0.87 - 1.18$\,ns, and $\tau_\text{In,ZPL} \sim 1.14 - 1.47$\,ns. 
The corresponding oscillator strengths are $f_\text{Al} \sim 0.42$, $f_\text{Ga} \sim 0.38$, and $f_\text{In} \sim 0.32$. 

The refractive index used in Eq.~\ref{eq:osc_strength} is determined from bulk transmission measurements presented in Ref.~\cite{niaouris2023col}, using the analytical solutions provided in Ref.~\cite{nichelatti2002aed}, yielding $n\left(\omega_{\text{Al}}\right) \simeq n\left(\omega_{\text{Ga}}\right) \simeq 2.9$, and $n\left(\omega_{\text{In}}\right) \simeq 2.7$.

\section{SIMS measurement of Al and Ga concentration in ZnO substrate}
\label{si:donor_conc}

\begin{figure}[!h]
    \centering
    \includegraphics[width=0.6\linewidth]{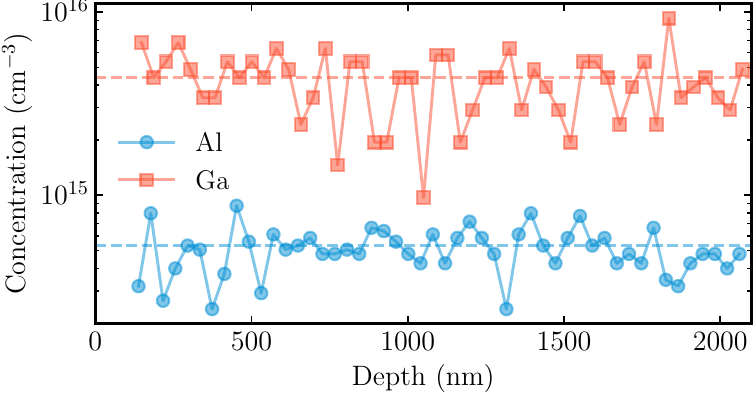}
    \caption{\label{fig:donor_conc}
    Donor concentration as a function of depth for the first two micrometers of the back surface of the parent substrate.
    }
\end{figure}

The Al and Ga donor concentration in the parent substrate was measured via secondary ion mass spectrometry (SIMS) measurements  on the back surface of the substrate. 
As depicted in Fig.~\ref{fig:donor_conc}, a relatively uniform donor density is observed through the two-micron measurement depth with an average density of  $5.3\cdot10^{14}$\,\cmnV\ for Al and $4.4\cdot10^{15}$\,\cmnV\ for Ga.

\section{Experimental Setup and Equipment}
\label{si:microscope}

\begin{figure}[h]
  \centering
  \includegraphics[width=0.6\linewidth]{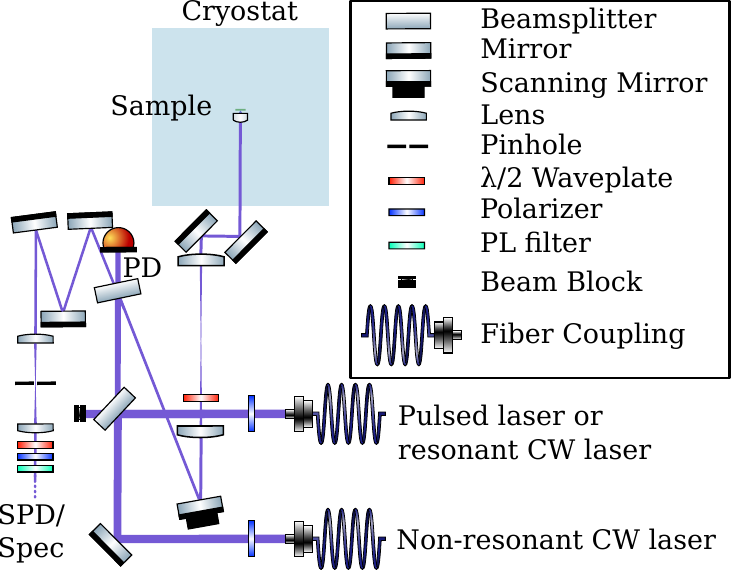}
  \caption{
  \label{fig:opt_path} 
  Optical path for the micro-PL/PLE experiments. A reference to the incident laser power is picked-off by a photodiode (PD). This experiment utilizes an aspheric 3.1\,mm-focus 0.62 NA lens inside the sample space for PL collection. The emitted PL is detected with either a single photon detector (SPD) or a spectrometer (Spec).
  }
\end{figure}

Fig.~\ref{fig:opt_path} depicts the optical path of the experiments described in the main text. 

Due to the high refractive index of ZnO at the In-donor emission wavelength relative to the superfluid He and He gas environment ($\mathrm{n_{\mathrm{ZnO}}({\lambda_{In}}}) = 2.7 > \mathrm{n}_\mathrm{He}({\lambda_{In}}) = 1$), only $\sim 1$\,\% of emitted photons are collected by the first 0.62 NA aspheric lens. 
The downstream microscope efficiency is $3$\,\% including the $35$\,\% detection efficiency, resulting in an overall system efficiency of $\sim 0.03$\,\%.
With this $\sim 0.03$\,\% efficiency, we estimate an overall PL emission rate of $\sim 5\times10^7$ photons/second assuming a collection rate of 14,000 photons/second (see \ref{fig:donor_ver}b), which was measured below the saturation limit.

\section{Lamella preparation process}
\label{si:sample_prep}

\begin{figure}[!h]
    \centering
    \includegraphics[width=0.50\linewidth]{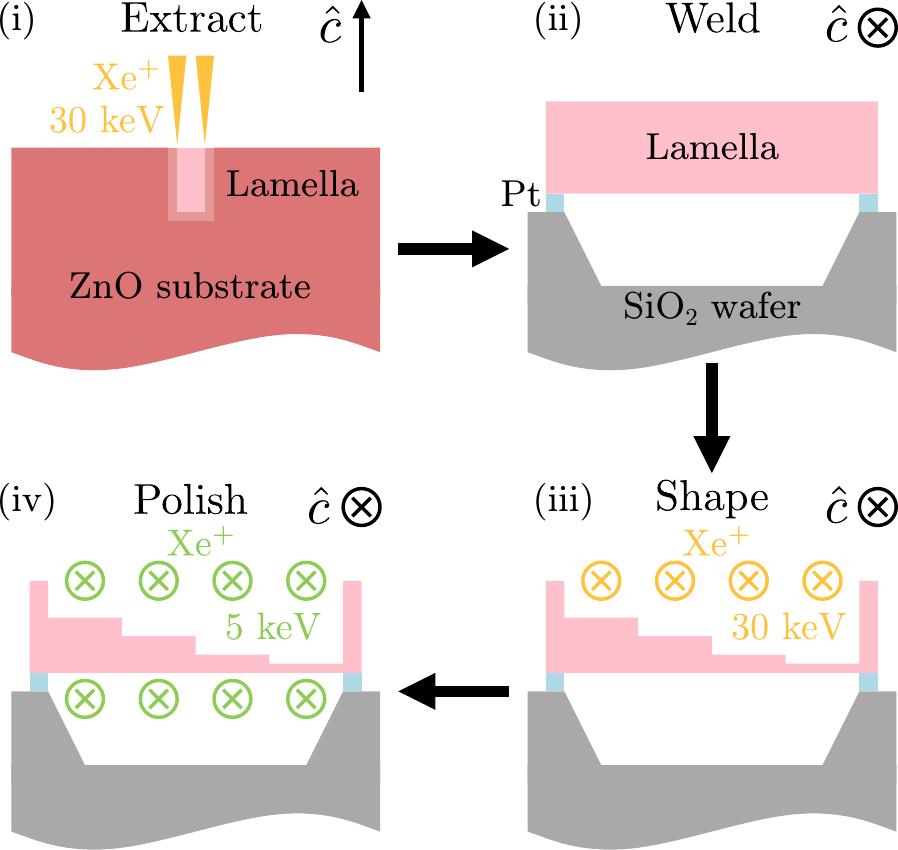}
    \caption{\label{fig:sample_prep}
    Lamella preparation process: (i) lamella extraction from ZnO substrate, (ii) lamella welded on Si\Otwo\ wafer, (iii) PFIB milling of lamella, (iv) removal of high energy PFIB-related damage with low energy PFIB.
    }
\end{figure}

In the main text, we describe the lamella preparation process. 
Fig.~\ref{fig:sample_prep} illustrates this process through steps i--iv. 

\section{Spectral kinetic series PL measurements before and after annealing}
\label{si:donor_stability}

\begin{figure}[!h]
    \centering
    \includegraphics[width=0.99\linewidth]{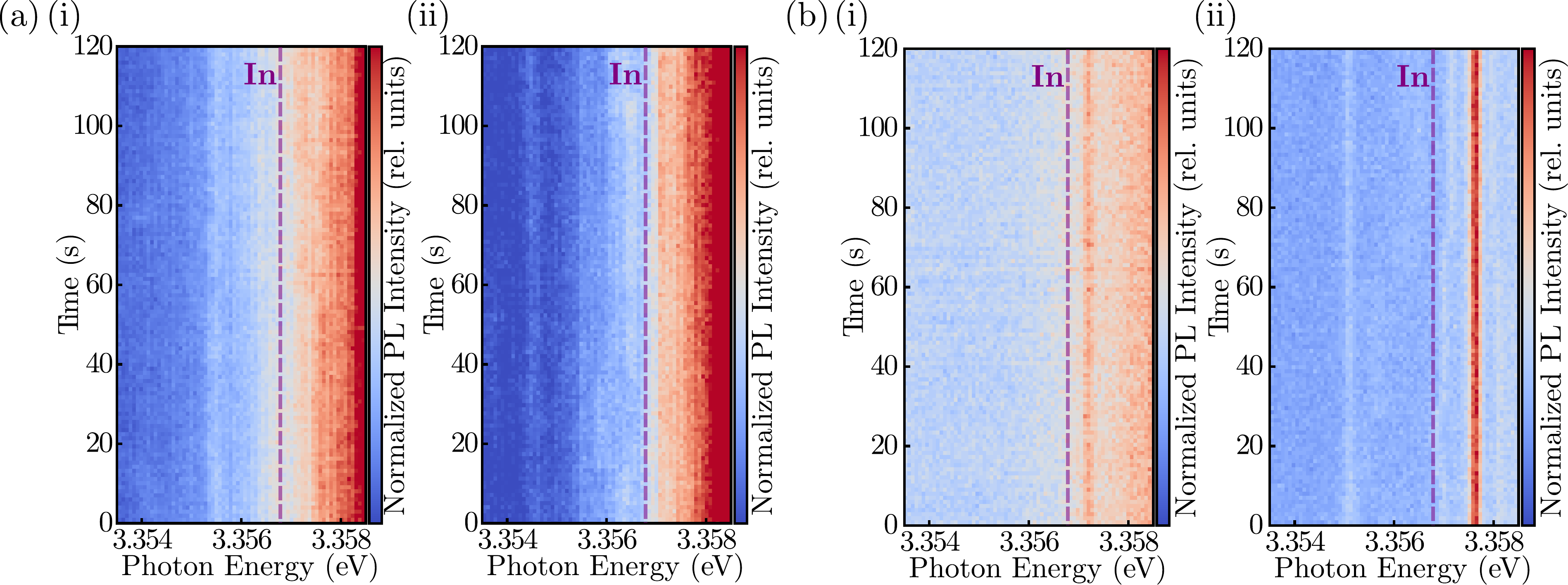}
    \caption{\label{fig:In_kinetic_series}
    PL spectra collected with a 1 second exposure time. Excitation is 3.44\,eV.
    (a) Spectra collected from the pre-annealed lamella within region B (i) and region C (ii).
    (b) Spectra collected from the annealed lamella within region C (i) and region D (ii). }
\end{figure}

Fig.~\ref{fig:In_kinetic_series}a shows kinetic series from the pre-annealed lamella at a location in region B (i) and region C (ii).
Similar to Fig.~\ref*{fig:spectra}c in the main text, we observe several weak peaks in the \InX\ region which spectrally diffuse over time. 

Fig.~\ref{fig:In_kinetic_series}b shows kinetic series at a location in region C (i) and region D (ii) after annealing. 
Similar to Fig.~\ref*{fig:spectra}e in the main text, stronger lines near the \InX\ energy are observed that are spectrally stable in time. 
While the luminescence is dramatically improved with annealing, as shown in Fig.~\ref{fig:In_kinetic_series}b-ii, at some locations weak emission and spectral diffusion can still be observed.

\section{Single Emitter PL spectrum, uncorrected}
\label{si:emitter_PL}

\begin{figure}[!h]
    \centering
    \includegraphics[width=0.6\linewidth]{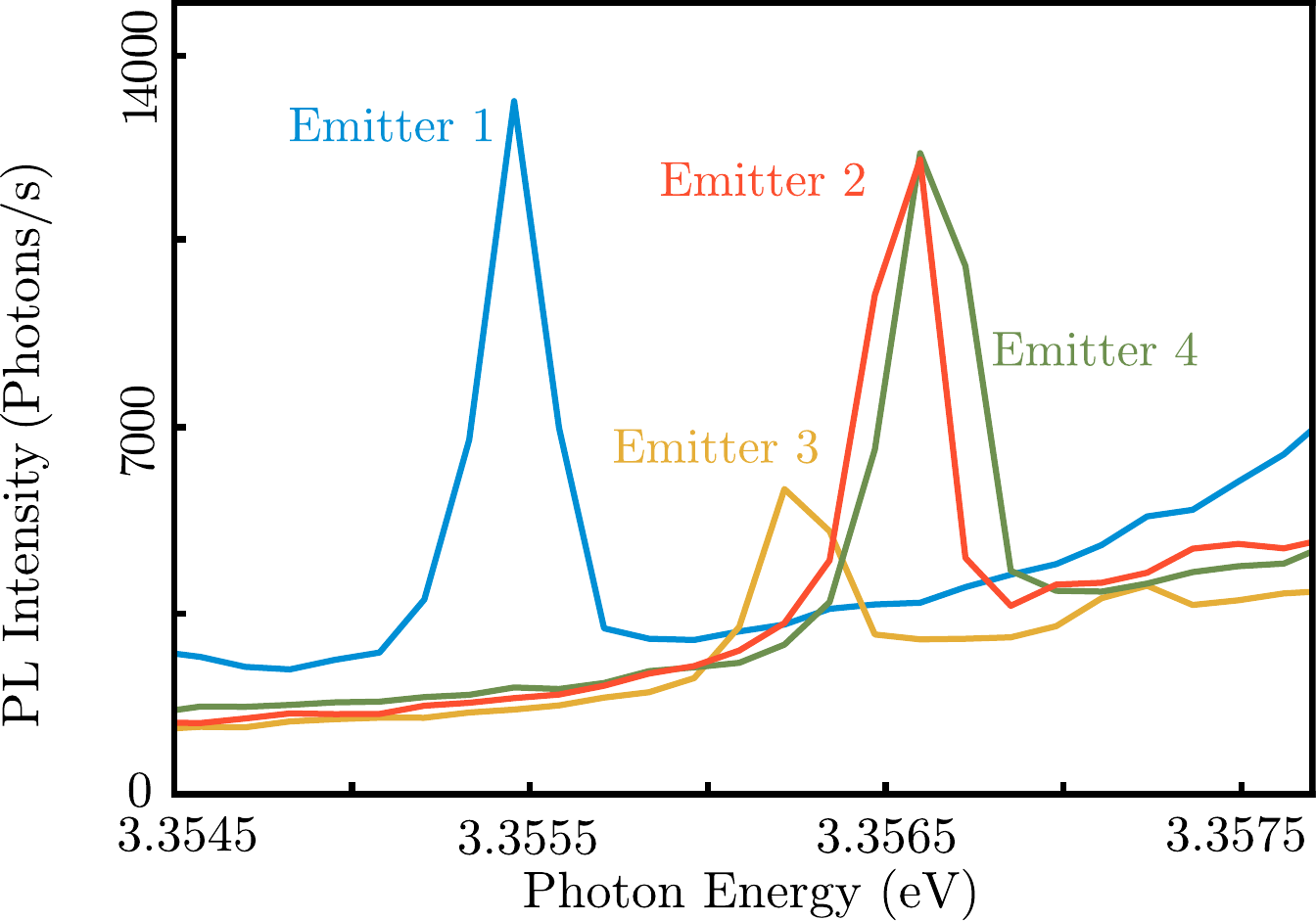}
    \caption{\label{fig:raw_In_PL}
    PL spectra of emitters 1--4 at 5.2\,K. Excitation is 3.44\,eV.
    }
\end{figure}

In Fig.~\ref*{fig:donor_ver}b of the main text, each single emitter PL spectra is shifted utilizing the location of the \AlX\ and \Yo\ lines to adjust for relative energy offsets due to strain.
As a result of the correction, the three emitters identified as In donors (emitters $2-4$) lie at the same energy within the spectrometer resolution, however emitter 1 still has exhibits a lower transition energy.
For completion, Fig.~\ref{fig:raw_In_PL} shows the uncorrected PL spectra. 

\section{Resonant sideband emission}
\label{si:res_tes}

\begin{figure}[!h]
    \centering
    \includegraphics[width=1.0\linewidth]{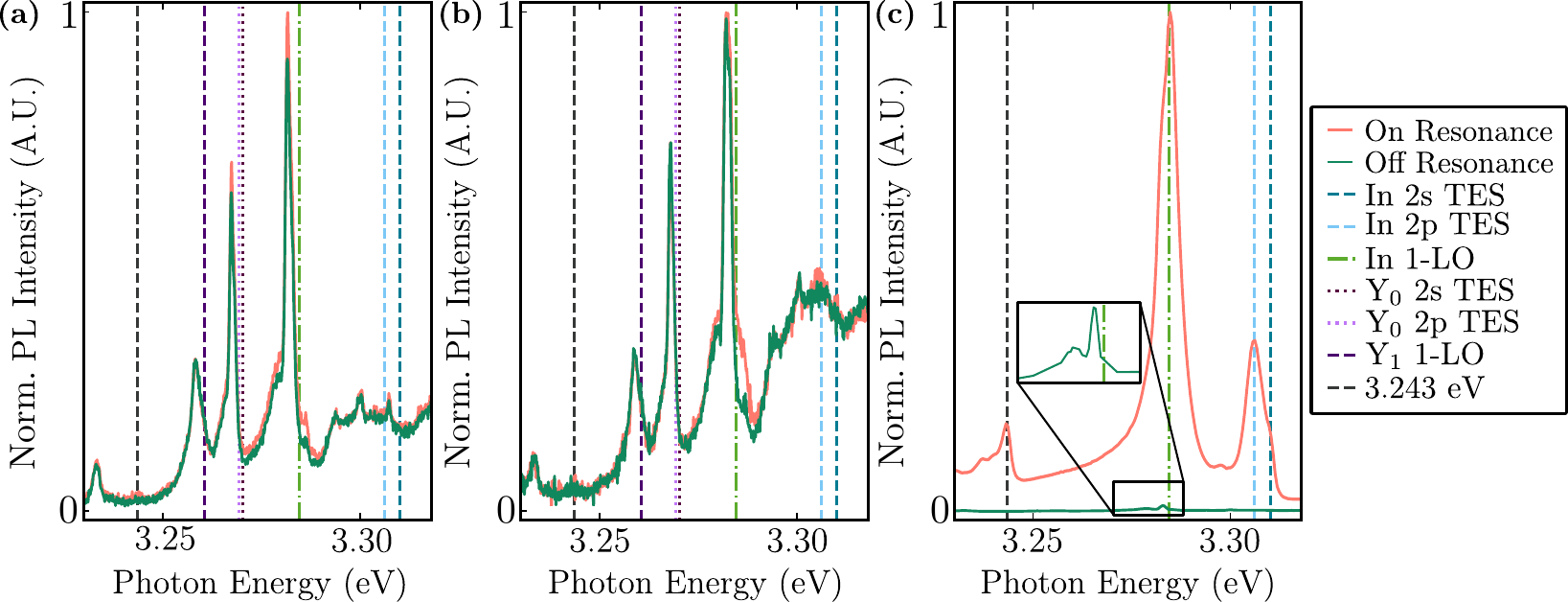}
    \caption{\label{fig:tes_em3_em4_implnt}
    Sideband PL with resonant and off-resonant excitation, normalized to the on-resonance spectrum}. Reported transitions~\cite{wagner2011bez} are marked with vertical dashed lines. 
    (a) Emitter 3 sideband PL with 3.356462\,eV resonant and 3.56417\,eV off-resonant excitation. 
    (b) Emitter 4 sideband PL with 3.356706\,eV resonant and 3.356681\,eV off-resonant excitation. 
    (c) Sideband PL for implanted In sample with 3.357244\,eV resonant and 3.356944\,eV off-resonant excitation~\cite{wang2023pdq}.
\end{figure}

Fig.~\ref*{fig:donor_ver}c in the main text depicts the sideband PL of emitter 2 under resonant \InX\ excitation. 
For completeness, Fig.~\ref{fig:tes_em3_em4_implnt}a and Fig.~\ref{fig:tes_em3_em4_implnt}b shows the observed sideband PL for emitters 3 and 4, respectively. 
Fig.~\ref{fig:tes_em3_em4_implnt}c shows PL sideband from a similar ZnO substrate implanted with In donors~\cite{wang2023pdq}, correlating the 3.243\,eV sideband feature observed in Fig.~\ref*{fig:donor_ver}c in the main text and Figs.~\ref{fig:tes_em3_em4_implnt}a and~\ref{fig:tes_em3_em4_implnt}b with an In transition.
Additionally, under off-resonant excitation, the background near the \In\ 1-LO peak in this In-implanted sample (Fig.~\ref{fig:tes_em3_em4_implnt}c inset) is similar to the background observed on the sideband spectra of emitters 2, 3, and 4.

\section{Polarization selection rules}
\label{si:pol_rules}

\begin{figure}[!h]
    \centering
    \includegraphics[width=1.0\linewidth]{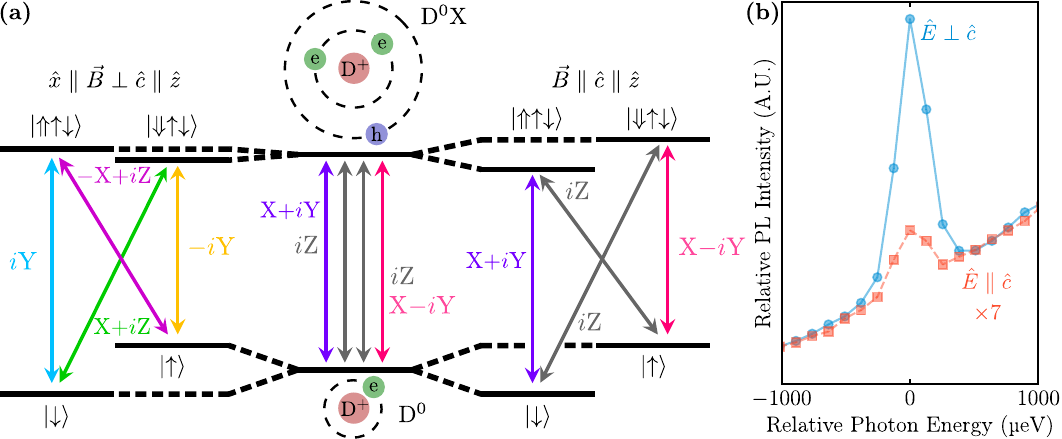}
    \caption{\label{fig:pol_rules}
    (a) Polarization selection rules with $\vec{B}\parallel\hat{x}$ (left), $B=0$ (middle) and $\vec{B}\parallel\hat{z}$ (right) in the sample reference frame ($\hat{c}\parallel\hat{z}$).
    Here, \up\ and \down\ (\uph\ and \downh) denote the electron (hole) spin up and down states.
    $\text{X} = \frac{a}{\sqrt{2}}\hat{x}$, $\text{Y} = \frac{a}{\sqrt{2}}\hat{y}$, and $\text{Z} = b\hat{z}$~\cite{linpeng2020dqd}, where $a=0.995$ and $b=0.0999$~\cite{lambrecht2002vbo}.
    (b) O-field PL of emitter 2, collecting polarization parallel and perpendicular to the $\hat c$ axis. 
    }
\end{figure}

PL emission polarization depends on the applied magnetic field orientation in relation to the crystal axis $\hat{c}$. 
Following Ref.~\cite{wagner2009vbs,lambrecht2002vbo,linpeng2020dqd}, Fig.~\ref{fig:pol_rules}a shows the expected photon polarization selection rules for the different transitions and magnetic field geometries, when $\hat{c}\parallel\hat{z}$.

In our experimental setup, the optical axis is always parallel to the $\hat{y}$ axis.
Hence, we cannot detect transitions with polarization components parallel to $\hat{y}$ (where $\hat{y}\perp\hat{c}$).
Therefore, we expect the PL emission polarized perpendicular to the crystal axis ($\hat{E}\parallel\hat{x}\perp\hat{c}$) to be $\sim 50\times$ brighter than emission polarized parallel to the crystal axis ($\hat{E}\parallel\hat{z}\parallel\hat{c}$).
Fig.~\ref{fig:pol_rules}b shows the emission at the two different polarizations at 0\,T for emitter 2. 
We are able to observe the weak $\hat{E}\parallel\hat{c}$ emission, and find that \InX\ $\hat{E}\perp\hat{c}$ emission is $\sim 28\times$ brighter than $\hat{E}\parallel\hat{c}$, which is close to the theoretical $50$.
Fig.~\ref{fig:pol_rules}a also depicts the polarization selection rules for \Bparc.
In this geometry, we were unable to detect the weak $\hat{E}\parallel\hat{c}$ peaks.

\section{Magneto-PL}
\label{si:magnetoPL}

\begin{figure}[!h]
    \centering
    \includegraphics[width=0.7\linewidth]{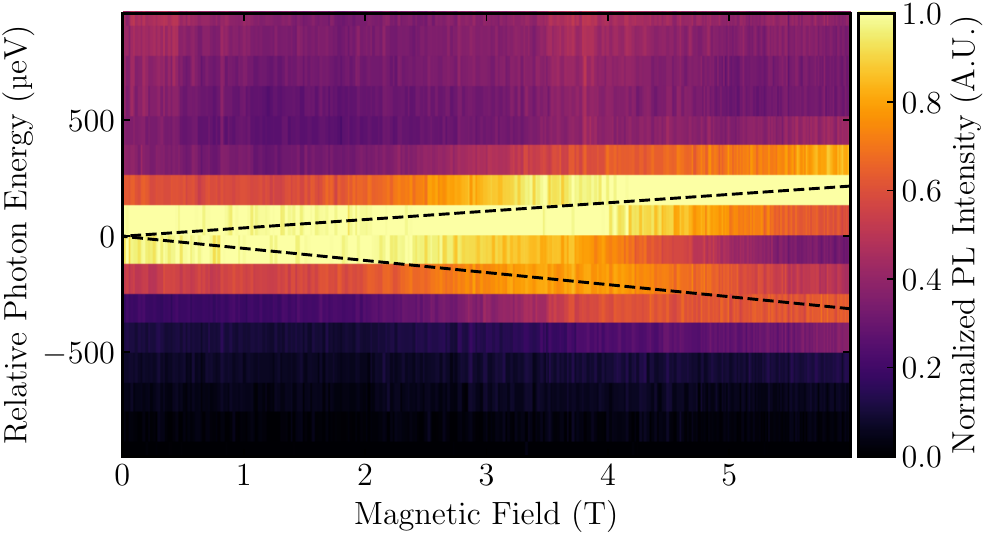}
    \caption{\label{fig:magnetoPL}
    PL spectra as a function of magnetic field on emitter 4 in \Bperpc\ orientation. Each PL spectrum (single vertical line) is normalized to the maximum PL counts in the depicted energy region after subtracting the minimum PL counts in the same region. The dashed lines correspond to a linear fit of the centers of two Gaussian functions, yielding $g_{\text{eff}} = 1.52 \pm 0.01$.
    }
\end{figure}

Figures \ref*{fig:donor_ver}d and \ref*{fig:donor_ver}e of the main text depict PL spectra of the four emitters at 0\,T and 6\,T. 
Fig.~\ref{fig:magnetoPL} depicts PL spectra as a function of magnetic field for emitter 4 in the \Bperpc\ orientation.
\section{Lifetime measurement background}
\label{si:lifetime_bg}

\begin{figure}[!h]
    \centering
    \includegraphics[width=0.55\linewidth]{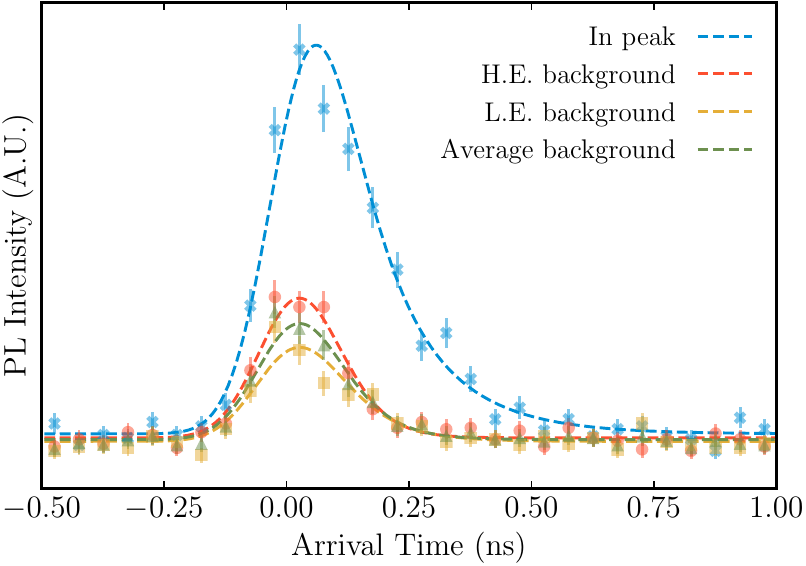}
    \caption{\label{fig:lifetime_bg}
    Example lifetime measurement on emitter 3, depicting the measured lifetime on the emitter peak, the Al background on the higher (H.E.) and lower (L.E.) energy tails ($\sim \pm 60$\,GHz), and the averaged of the two background contributions.
    }
\end{figure}

When performing time-resolved measurements on the In emitter, part of the collected photoluminescence to corresponds to background emission; the single In emitters lie on the tail of the Al/Ga emission.
To estimate the background contribution, we perform lifetime measurements collecting PL from the high and low energy edges ($\sim \pm 60$\,GHz) of the In emission. 
Assuming the background contribution is nearly linear, we average the two lifetime background data and fit to an exponential-Gaussian convolution profile.
We use the fitted background parameters as fixed parameters to the double exponential-Gaussian profile fitted to the In emitter lifetime data (Fig.~\ref{fig:lifetime_bg}). 
For the data presented in Fig.~\ref{fig:lifetime_bg}, the background lifetime is $110 \pm 20$\,ps.

\section{Temperature dependence of \texorpdfstring{\InX}~~PL intensity}
\label{si:temp_dep}

\begin{figure}[h]
    \centering
    \includegraphics[width=0.55\linewidth]{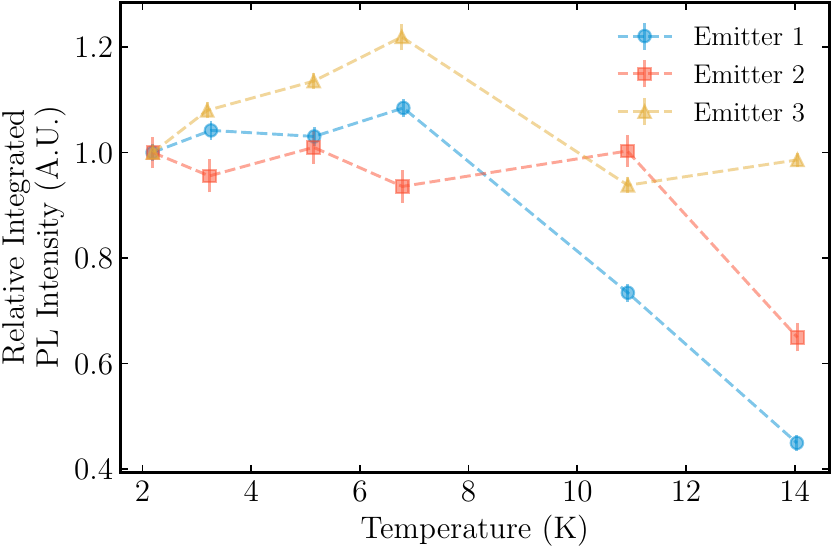}
    \caption{\label{fig:temp_dep_PL}
    Temperature dependence of the integrated PL intensity for emitters 1, 2, and 3.
    For each emitter we normalize all data to their lowest temperature values. 
    Excitation is 3.44\,eV.
    }
\end{figure}

To investigate whether the non-radiative recombination rate is affected by temperature, we measure the PL intensity of the \InX\ transition as a function of temperature.
Fig.~\ref{fig:temp_dep_PL} shows that the PL intensity is fairly constant between 2\,K and 6\,K, indicating that the lifetime is constant throughout this temperature range.

\bibliography{main}